%
%
\def\unredoffs{} \def\redoffs{\voffset=-.31truein\hoffset=-.59truein}
\def\speclscape{}
%
%
%
%
\newbox\leftpage \newdimen\fullhsize \newdimen\hstitle \newdimen\hsbody
\tolerance=1000\hfuzz=2pt
\catcode`\@=11 
\def\bigans{b }
\def\answ{b }
%

\ifx\answ\bigans\message{(This will come out unreduced.}
\magnification=1200\unredoffs\baselineskip=.33truein plus 2pt minus 1pt
\hsbody=\hsize \hstitle=\hsize 
\else\message{(This will be reduced.} \let\l@r=L
\magnification=1000\baselineskip=16pt plus 2pt minus 1pt \vsize=7truein
\redoffs \hstitle=8truein\hsbody=4.75truein\fullhsize=10truein\hsize=\hsbody
\output={\ifnum\pageno=0 
  \shipout\vbox{\speclscape{\hsize\fullhsize\makeheadline}
   \hbox to \fullhsize{\hfill\pagebody\hfill}}\advancepageno
  \else
 \almostshipout{\leftline{\vbox{\pagebody\makefootline}}}\advancepageno
  \fi}
\def\almostshipout#1{\if L\l@r \count1=1 \message{[\the\count0.\the\count1]}
      \global\setbox\leftpage=#1 \global\let\l@r=R
 \else \count1=2
  \shipout\vbox{\speclscape{\hsize\fullhsize\makeheadline}
      \hbox to\fullhsize{\box\leftpage\hfil#1}}  \global\let\l@r=L\fi}
\fi
%
\newcount\yearltd\yearltd=\year\advance\yearltd by -1900

\def\Title#1#2{\nopagenumbers\abstractfont\hsize=\hstitle\rightline{#1}%
\vskip 1in\centerline{\titlefont #2}\abstractfont\vskip .5in\pageno=0}
\def\Date#1{\vfill\leftline{#1}\tenpoint\supereject\global\hsize=\hsbody%
\footline={\hss\tenrm\folio\hss}}
%

\def\draftmode{\message{ DRAFTMODE }\def\draftdate{{\rm preliminary draft:
\number\month/\number\day/\number\yearltd\ \ \hourmin}}%
\headline={\hfil\draftdate}\writelabels\baselineskip=20pt plus 2pt minus 2pt
 {\count255=\time\divide\count255 by 60 \xdef\hourmin{\number\count255}
  \multiply\count255 by-60\advance\count255 by\time
  \xdef\hourmin{\hourmin:\ifnum\count255<10 0\fi\the\count255}}}
\def\nolabels{\def\wrlabeL##1{}\def\eqlabeL##1{}\def\reflabeL##1{}}
\def\writelabels{\def\wrlabeL##1{\leavevmode\vadjust{\rlap{\smash%
{\line{{\escapechar=` \hfill\rlap{\sevenrm\hskip.03in\string##1}}}}}}}%
\def\eqlabeL##1{{\escapechar-1\rlap{\sevenrm\hskip.05in\string##1}}}%
\def\reflabeL##1{\noexpand\llap{\noexpand\sevenrm\string\string\string##1}}}
\nolabels
%
\global\newcount\secno \global\secno=0
\global\newcount\meqno \global\meqno=1
\def\newsec#1{\global\advance\secno by1\message{(\the\secno. #1)}

\global\subsecno=0\eqnres@t\noindent{\bf\the\secno. #1}
\writetoca{{\secsym} {#1}}\par\nobreak\medskip\nobreak}
\def\eqnres@t{\xdef\secsym{\the\secno.}\global\meqno=1\bigbreak\bigskip}
\def\sequentialequations{\def\eqnres@t{\bigbreak}}\xdef\secsym{}
\global\newcount\subsecno \global\subsecno=0
\def\subsec#1{\global\advance\subsecno by1\message{(\secsym\the\subsecno. #1)}
\ifnum\lastpenalty>9000\else\bigbreak\fi
\noindent{\sl\secsym\the\subsecno. #1}\writetoca{\string\quad
{\secsym\the\subsecno.} {#1}}\par\nobreak\medskip\nobreak}

\def\appendix#1#2{\global\meqno=1\global\subsecno=0\xdef\secsym{\hbox{#1.}}
\bigbreak\bigskip\noindent{\bf Appendix #1. #2}\message{(#1. #2)}
\writetoca{Appendix {#1.} {#2}}\par\nobreak\medskip\nobreak}
%
%
\def\eqnn#1{\xdef #1{(\secsym\the\meqno)}\writedef{#1\leftbracket#1}%
\global\advance\meqno by1\wrlabeL#1}
\def\eqna#1{\xdef #1##1{\hbox{$(\secsym\the\meqno##1)$}}
\writedef{#1\numbersign1\leftbracket#1{\numbersign1}}%
\global\advance\meqno by1\wrlabeL{#1$\{\}$}}
\def\eqn#1#2{\xdef #1{(\secsym\the\meqno)}\writedef{#1\leftbracket#1}%
\global\advance\meqno by1$$#2\eqno#1\eqlabeL#1$$}
%
\newskip\footskip\footskip14pt plus 1pt minus 1pt 
\def\footnotefont{\ninepoint}\def\f@t#1{\footnotefont #1\@foot}
\def\f@@t{\baselineskip\footskip\bgroup\footnotefont\aftergroup\@foot\let\next}
\setbox\strutbox=\hbox{\vrule height9.5pt depth4.5pt width0pt}
\global\newcount\ftno \global\ftno=0
\def\foot{\global\advance\ftno by1\footnote{$^{\the\ftno}$}}
%
\newwrite\ftfile
\def\footend{\def\foot{\global\advance\ftno by1\chardef\wfile=\ftfile
$^{\the\ftno}$\ifnum\ftno=1\immediate\openout\ftfile=foots.tmp\fi%
\immediate\write\ftfile{\noexpand\smallskip%
\noexpand\item{f\the\ftno:\ }\pctsign}\findarg}%
\def\footatend{\vfill\eject\immediate\closeout\ftfile{\parindent=20pt
\centerline{\bf Footnotes}\nobreak\bigskip\input foots.tmp }}}
\def\footatend{}
%
%
\global\newcount\refno \global\refno=1
\newwrite\rfile
\def\ref{[\the\refno]\nref}
\def\nref#1{\xdef#1{[\the\refno]}\writedef{#1\leftbracket#1}%
\ifnum\refno=1\immediate\openout\rfile=refs.tmp\fi
\global\advance\refno by1\chardef\wfile=\rfile\immediate
\write\rfile{\noexpand\item{#1\ }\reflabeL{#1\hskip.31in}\pctsign}\findarg}
\def\findarg#1#{\begingroup\obeylines\newlinechar=`\^^M\pass@rg}
{\obeylines\gdef\pass@rg#1{\writ@line\relax #1^^M\hbox{}^^M}%
\gdef\writ@line#1^^M{\expandafter\toks0\expandafter{\striprel@x #1}%
\edef\next{\the\toks0}\ifx\next\em@rk\let\next=\endgroup\else\ifx\next\empty%
\else\immediate\write\wfile{\the\toks0}\fi\let\next=\writ@line\fi\next\relax}}
\def\striprel@x#1{} \def\em@rk{\hbox{}}
\def\lref{\begingroup\obeylines\lr@f}
\def\lr@f#1#2{\gdef#1{\ref#1{#2}}\endgroup\unskip}

\def\addref#1{\immediate\write\rfile{\noexpand\item{}#1}} 
\def\startrefs#1{\immediate\openout\rfile=refs.tmp\refno=#1}
\def\xref{\expandafter\xr@f}\def\xr@f[#1]{#1}
\def\refs#1{\count255=1[\r@fs #1{\hbox{}}]}
\def\r@fs#1{\ifx\und@fined#1\message{reflabel \string#1 is undefined.}%
\nref#1{need to supply reference \string#1.}\fi%
\vphantom{\hphantom{#1}}\edef\next{#1}\ifx\next\em@rk\def\next{}%
\else\ifx\next#1\ifodd\count255\relax\xref#1\count255=0\fi%
\else#1\count255=1\fi\let\next=\r@fs\fi\next}
%

%
\newwrite\ffile\global\newcount\figno \global\figno=1
\def\fig{Figure~\the\figno\nfig}
\def\nfig#1{\xdef#1{Figure~\the\figno}%
\writedef{#1\leftbracket fig.\noexpand~\the\figno}%
\ifnum\figno=1\immediate\openout\ffile=figs.tmp\fi\chardef\wfile=\ffile%
\immediate\write\ffile{\noexpand\medskip\noexpand\item{Fig.\ \the\figno. }
\reflabeL{#1\hskip.55in}\pctsign}\global\advance\figno by1\findarg}
\def\lfig{\begingroup\obeylines\lf@g}
\def\lf@g#1#2{\gdef#1{\fig#1{#2}}\endgroup\unskip}
\def\xfig{\expandafter\xf@g}\def\xf@g fig.\penalty\@M\ {}
\def\figs#1{figs.~\f@gs #1{\hbox{}}}
\def\f@gs#1{\edef\next{#1}\ifx\next\em@rk\def\next{}\else
\ifx\next#1\xfig #1\else#1\fi\let\next=\f@gs\fi\next}
\newwrite\lfile
{\escapechar-1\xdef\pctsign{\string\%}\xdef\leftbracket{\string\{}
\xdef\rightbracket{\string\}}\xdef\numbersign{\string\#}}

\def\writestop{\def\writestoppt{\immediate\write\lfile{\string\pageno%
\the\pageno\string\startrefs\leftbracket\the\refno\rightbracket%
\string\def\string\secsym\leftbracket\secsym\rightbracket%
\string\secno\the\secno\string\meqno\the\meqno}\immediate\closeout\lfile}}
\def\writestoppt{}\def\writedef#1{}
\def\seclab#1{\xdef #1{\the\secno}\writedef{#1\leftbracket#1}\wrlabeL{#1=#1}}
\def\subseclab#1{\xdef #1{\secsym\the\subsecno}%
\writedef{#1\leftbracket#1}\wrlabeL{#1=#1}}
\newwrite\tfile \def\writetoca#1{}
\def\leaderfill{\leaders\hbox to 1em{\hss.\hss}\hfill}
\def\writetoc{\immediate\openout\tfile=toc.tmp
   \def\writetoca##1{{\edef\next{\write\tfile{\noindent ##1
   \string\leaderfill {\noexpand\number\pageno} \par}}\next}}}

%
\catcode`\@=12 
%
\edef\tfontsize{\ifx\answ\bigans scaled\magstep3\else scaled\magstep4\fi}
\font\titlerm=cmr10 \tfontsize \font\titlerms=cmr7 \tfontsize
\font\titlermss=cmr5 \tfontsize \font\titlei=cmmi10 \tfontsize
\font\titleis=cmmi7 \tfontsize \font\titleiss=cmmi5 \tfontsize
\font\titlesy=cmsy10 \tfontsize \font\titlesys=cmsy7 \tfontsize
\font\titlesyss=cmsy5 \tfontsize \font\titleit=cmti10 \tfontsize
\skewchar\titlei='177 \skewchar\titleis='177 \skewchar\titleiss='177
\skewchar\titlesy='60 \skewchar\titlesys='60 \skewchar\titlesyss='60
\def\titlefont{\def\rm{\fam0\titlerm}
\textfont0=\titlerm \scriptfont0=\titlerms \scriptscriptfont0=\titlermss
\textfont1=\titlei \scriptfont1=\titleis \scriptscriptfont1=\titleiss
\textfont2=\titlesy \scriptfont2=\titlesys \scriptscriptfont2=\titlesyss
\textfont\itfam=\titleit \def\it{\fam\itfam\titleit}\rm}
 \ifx\answ\bigans\else scaled\magstep1\fi
\ifx\answ\bigans\def\abstractfont{\tenpoint}\else
\font\abssl=cmsl10 scaled \magstep1
\font\absrm=cmr10 scaled\magstep1 \font\absrms=cmr7 scaled\magstep1
\font\absrmss=cmr5 scaled\magstep1 \font\absi=cmmi10 scaled\magstep1
\font\absis=cmmi7 scaled\magstep1 \font\absiss=cmmi5 scaled\magstep1
\font\abssy=cmsy10 scaled\magstep1 \font\abssys=cmsy7 scaled\magstep1
\font\abssyss=cmsy5 scaled\magstep1 \font\absbf=cmbx10 scaled\magstep1
\skewchar\absi='177 \skewchar\absis='177 \skewchar\absiss='177
\skewchar\abssy='60 \skewchar\abssys='60 \skewchar\abssyss='60
\def\abstractfont{\def\rm{\fam0\absrm}
\textfont0=\absrm \scriptfont0=\absrms \scriptscriptfont0=\absrmss
\textfont1=\absi \scriptfont1=\absis \scriptscriptfont1=\absiss
\textfont2=\abssy \scriptfont2=\abssys \scriptscriptfont2=\abssyss
\textfont\itfam=\bigit \def\it{\fam\itfam\bigit}\def\footnotefont{\tenpoint}%
\textfont\slfam=\abssl \def\sl{\fam\slfam\abssl}%
\textfont\bffam=\absbf \def\bf{\fam\bffam\absbf}\rm}\fi
\def\tenpoint{\def\rm{\fam0\tenrm}
\textfont0=\tenrm \scriptfont0=\sevenrm \scriptscriptfont0=\fiverm
\textfont1=\teni  \scriptfont1=\seveni  \scriptscriptfont1=\fivei
\textfont2=\tensy \scriptfont2=\sevensy \scriptscriptfont2=\fivesy
\textfont\itfam=\tenit \def\it{\fam\itfam\tenit}\def\footnotefont{\ninepoint}%
\textfont\bffam=\tenbf \def\bf{\fam\bffam\tenbf}\def\sl{\fam\slfam\tensl}\rm}
\font\ninerm=cmr9 \font\sixrm=cmr6 \font\ninei=cmmi9 \font\sixi=cmmi6
\font\ninesy=cmsy9 \font\sixsy=cmsy6 \font\ninebf=cmbx9
\font\nineit=cmti9 \font\ninesl=cmsl9 \skewchar\ninei='177
\skewchar\sixi='177 \skewchar\ninesy='60 \skewchar\sixsy='60
\def\ninepoint{\def\rm{\fam0\ninerm}
\textfont0=\ninerm \scriptfont0=\sixrm \scriptscriptfont0=\fiverm
\textfont1=\ninei \scriptfont1=\sixi \scriptscriptfont1=\fivei
\textfont2=\ninesy \scriptfont2=\sixsy \scriptscriptfont2=\fivesy
\textfont\itfam=\ninei \def\it{\fam\itfam\nineit}\def\sl{\fam\slfam\ninesl}%
\textfont\bffam=\ninebf \def\bf{\fam\bffam\ninebf}\rm}
%
%

\hyphenation{anom-aly anom-alies coun-ter-term coun-ter-terms}
\def\inv{^{\raise.15ex\hbox{${\scriptscriptstyle -}$}\kern-.05em 1}}

\def\Dsl{\,\raise.15ex\hbox{/}\mkern-13.5mu D} 
\def\dsl{\raise.15ex\hbox{/}\kern-.57em\partial}

\font\bigit=cmti10 scaled \magstep1
\def\lspace{\ifx\answ\bigans{}\else\qquad\fi}
\def\lbspace{\ifx\answ\bigans{}\else\hskip-.2in\fi} 
\def\boxeqn#1{\vcenter{\vbox{\hrule\hbox{\vrule\kern3pt\vbox{\kern3pt
	\hbox{${\displaystyle #1}$}\kern3pt}\kern3pt\vrule}\hrule}}}
\def\mbox#1#2{\vcenter{\hrule \hbox{\vrule height#2in
		\kern#1in \vrule} \hrule}}  
%

\def\grad#1{\,\nabla\!_{{#1}}\,}

\def\darr#1{\raise1.5ex\hbox{$\leftrightarrow$}\mkern-16.5mu #1}

\def\half{{\textstyle{1\over2}}} 
\def\roughly#1{\raise.3ex\hbox{$#1$\kern-.75em\lower1ex\hbox{$\sim$}}}

\def\perpp{{\!\scriptscriptstyle\perp}}

\def\half{{1\over 2}}

\def\dnb{\delta{\bf n}}

\def\grad{\nabla_\perpp}
\def\dot{\!\cdot\!}
\def\cross{\!\times\!}

\def\bold#1{\setbox0=\hbox{$#1$}%
     \kern-.010em\copy0\kern-\wd0
     \kern.025em\copy0\kern-\wd0
     \kern-.020em\raise.0200em\box0 }

\lref\dg{P.G.~de Gennes, Solid State Commun. {\bf 14} (1973) 997.}
\lref\RL{S.R.~Renn and T.C.~Lubensky, Phys. Rev. A {\bf 38} (1988) 2132;
{\bf 41} (1990) 4392.}
\lref\DIS{J.~Goodby, M.A.~Waugh, S.M.~Stein. R.~Pindak, and J.S.~Patel,
Nature {\bf 337} (1988) 449; J. Am. Chem. Soc. {\bf 111} (1989) 8119;
G.~Strajer, R.~Pindak, M.A.~Waugh, J.W.~Goodby, and J.S.~Patel, Phys. Rev.
Lett.
{\bf 64} (1990) 13; K.J.~Ihn, J.A.N.~Zasadzinski, R.~Pindak, A.J.~Slaney, and
J.~Goodby,
Science {\bf 258} (1992) 275.}
\lref\NLstar{T.~Chan, C.W.~Garland and H.T.~Nguyen, Phys. Rev. E {\bf 52}
(1995) 5000;
L.~Navailles, C.W.~Garland and H.T.~Nguyen, J. Phys. II France {\bf 6} (1996)
1243.}
\lref\KL{R.D.~Kamien and T.C.~Lubensky, J. Phys. I France {\bf 3} (1993) 2131.}
\lref\dgp{P.G.~de Gennes and J.~Prost, {\sl The Physics of Liquid Crystals},
Second Edition, (Oxford University Press, New York, 1993).}
\lref\KN{R.D.~Kamien and D.R.~Nelson, Phys. Rev. Lett. {\bf 74} (1995) 2499;
Phys. Rev. E
{\bf 53} (1996) 650.}
\lref\HPP{G.A.~Hinshaw, Jr., R.G.~Petschek and R.A.~Pelcovits, Phys. Rev. Lett.
{\bf 60} (1988) 1864.}
\lref\LY{G.~Yan and T.C.~Lubensky, University of Pennsylvania Preprint (1996)
[cond-mat/9611066].}
\lref\Pansu{B.~Pansu, M.L.~Hui and H.T.~Nguyen, {\sl to appear in} J. Phys. II
France (1997).}
\lref\NS{D.R.~Nelson, Phys. Rev. Lett. {\bf60} (1988) 1973;
D.R.~Nelson and H.S. Seung, Phys. Rev. B {\bf 39} (1989) 9153.}
\lref\SWM{S.~Meiboom, J.P.~Sethna, P.W.~Anderson and W.F.~Brinkman, Phys. Rev.
Lett. {\bf 46} (1981) 1216;
D.C.~Wright and N.D.~Mermin, Rev. Mod. Phys. {\bf 61} (1989) 385.}

\nfig\fone{Double-twisted smectic structure.  The dark lines are
the screw dislocations.  Note that the layer normal
{\sl and} the screw dislocation tangent vectors have a double-twist texture.
While the rotation angles, defect density and layer spacing are related, this
figure does not take that into account -- it only serves as a simple model.}
\nfig\ftwo{A cut-away view of the double-twisted structure of \fone\
through the center of the tube.
}
\nfig\fsix{Geometrical construction used to derive \etanitanii\ and \eaiaii .
Coming out radially from the center the spacing of the layers
along the tube direction must be a constant $\delta$ in one
smectic slab.  Thus $R\tan\theta(R)=\delta$ and $\delta\cos\theta(R)=a(R)$
from which follow \etanitanii\ and \eaiaii , respectively.
}

\nfig\ffour{The $O^5$ blue phase structure.  The double-twist cylinders
are of radius $\rho$ and each side of the cube shown has length $4\rho$.
The three cylinders that are cut by the visible faces are labeled $A$, $B$ and
$C$,
while two of the kissing points are labeled $\Gamma$ and $\Delta$.}
\nfig\ffive{An unwrapped double-twist cylinder.  The plane
corresponds to the {\sl inside} surface of the cylinder $A$ in Figure 4.  The
dashed
line is that line parallel to the double-twist axis that is at an angle $\pi/2$
from
the similar line in which $\Gamma$ lies.  At $\Gamma$ the cylinder kisses
cylinder $B$, while at $\Delta$ it kisses cylinder $C$.
}
\Title{}{Smectic Order in Double-Twist Cylinders}
\centerline{
Randall D. Kamien\footnote{$^\dagger$}{kamien@dept.physics.upenn.edu}}
\smallskip
{\baselineskip=0.20truein\centerline{\sl Department of Physics and
Astronomy, University of Pennsylvania,}\centerline{
\sl Philadelphia, PA 19104, USA}}

\vskip .3in
I propose a double-twist texture with local smectic order, which
may have been seen in recent experiments \Pansu .  As in the
Renn-Lubensky TGB phase, the smectic order is broken only through a
lattice of screw dislocations.  A melted lattice of screw dislocations
can produce a double-twist texture as can an unmelted lattice.
In the latter case I show that geometry only allows for certain
angles between smectic regions.
I discuss the possibility of connecting these double-twist tubes
together to form a smectic blue phase.
\bigskip
\bigskip
\bigskip

\Date{4 November 1996; revised 13 February 1997}
\newsec{Introduction}
Since the prediction \RL\ and discovery \DIS\ of the defect-laden,
Renn-Lubensky, twist-grain-boundary ground state of a chiral smectic,
it has become {\sl de rigueur} to consider defected ground states in
liquid crystals \refs{\KN,\LY}.  While it is possible to introduce
defects into smectics or columnar phases to produce planar
chiral textures like the cholesteric or twisted hexatic \ref\KCL{R.D.~Kamien,
J. Phys. II France {\bf 6} (1996) 461.}, there has been no attempt to
construct the more complicated blue phase textures
\SWM\ in a similar fashion.
Recent experiments \Pansu, however, have suggested that there exist blue phases
with some sort of smectic order.  In this letter, I propose a general defect
structure that can produce smectic double-twist.  The smectic order may be only
{\sl short-ranged} as is seen in the $N_L^*$ phase of some liquid
crystals \refs{\NLstar,\KL}.  I will first discuss this case, which is likely
to be the generic case, and will then consider the possibility of having
long-range
smectic order.  In this case, the defects will form a regular lattice.  It will
be
shown that because of the periodicity imposed by the cylindrical structure of
a double-twist tube the allowed
rotation angles fall into a discrete class similar to, but not
as restrictive as, moir\'e angles \KN .

I start be recalling the general structure of a twist-grain-boundary in a
smectic liquid crystal.  The free energy for a smectic-$A$ can be written
to quadratic order
in terms of the layer displacement variable $u({\bf x})$ and the nematic
director
fluctuation $\delta{\bf n}({\bf x})={\bf n}-{\bf\hat z}$ \dg:
\eqn\efree{F=\half\int d^3\!x\,B(\partial_z u)^2 + B'(\nabla_\perpp u -
\delta{\bf n})^2
+K_1(\nabla_\perpp\dot\dnb)^2 + K_2(\grad\cross\dnb)^2 +
K_3(\partial_z\dnb)^2,}
where $B$ and $B'$ are bulk moduli and $K_i$ are the Frank elastic constants.
In the ground state, $\grad u=\dnb$, and so $\grad\cross\dnb\equiv 0$.  Thus
a smectic cannot have any twist.  However, as pointed out by Renn and Lubensky,
the smectic can twist if $u$ is not single-valued: in other words if there
are screw dislocations in the smectic structure \RL. Each screw dislocation has
a finite
energy per unit length and so there will be a balance between the cost of
a defect and the amount of twist produced by the structure.  Above some
critical chirality the defects will proliferate and organize themselves into a
sequence of twist-grain-boundaries.
In the twist-grain-boundary (TGB) state of
smectic-$A$ liquid crystals, the nematic director rotates around a pitch axis
$\bf\hat P$
which lies in the plane of the smectic layers.  The defected structure consists
of regions of perfect smectic layers, infinite in the two directions
perpendicular
to $\bf\hat P$ and of extent $d'$ along $\bf\hat P$.  Between each successive
slab there is a twist-grain-boundary consisting of parallel screw dislocations
with a uniform spacing $d$.  Each grain boundary effects a twist of the
director by $\Delta\theta = 2\sin^{-1}(b/2d)$ \ref\chalub{I thank
T. Lubensky for discussions on this point.  If you cannot arrange
an audience with him, see, for instance,
P.M.~Chaikin and T.C.~Lubensky, {\sl Principles of Condensed Matter Physics},
(Cambridge University Press, Cambridge, 1995), Chapter 9.}
where $b$ is the Burgers vector which
must be an integral multiple of the layer spacing $a$.
The lowest energy grain boundary for
a given rotation angle will consist of screw dislocations of unit strength,
{\sl i.e.}
$b=a$.  In the end, the TGB state has the director structure of a cholesteric
(with discretized jumps in angle) while at the same time having long-range
smectic order.

As temperature increases, if the smectic order does not melt first, the lattice
of screw dislocations can melt, in analogy with the melted flux-line liquid of
superconductors
\NS.   This $N_L^*$ phase \NLstar\ is thermodynamically the same as the
cholesteric
phase and
has a uniformly twisting director structure and is composed
of a melt of dislocation lines which themselves follow the director, as if they
were polymers in a polymer cholesteric phase \KL.  While there is no long-range
smectic order, there is short range smectic order which can be seen via X-ray
scattering
and through calorimetry: before the TGB state appears there is an absorption of
heat which is required to build up the local smectic order.

How can either of these dislocated structures be deformed to produce
double-twist?
Since the smectic order is locally preserved, there must be a defect structure
which produces an average double-twist, though possibly via non-uniform jumps.
Such a structure is shown in Figures 1 and 2, in two cross-sectional views.  In
this structure the layer normal has discretized double-twist as does the
tangent
vector field of the screw dislocations.  The structure of the double-twist
tube is identical to the usual TGB structure when {\sl wrapped} around a
central smectic region.  While the grain boundary
structure is highly distorted near the center
of the tube, it is clear that at larger radii the local structure at
a grain boundary is identical to the structure in the more usual
cholesteric-like
TGB state.  Thinking of the grain-boundaries as smectic layers, the grain
boundaries adopt a standard jelly-roll texture \dgp.  In the next section
I will discuss some of the geometric details required to build such a
structure.
Section 3 will discuss the energetics of the double-twist tube.  Finally, in
section 4, I will consider the possibility of a smectic blue phase.

\newsec{Geometric Details: Layer Spacing and Lock-In Angles}
While the above
geometric picture of the defect structure captures the essence of the
construction, the details are significantly more complex.  One
concern arises because the smectic layer tilt and spacing naturally change as
a function of radius.  In one undefected smectic region the spacing
between the layers along the tube axis ({\sl not} perpendicular to
the layers) must be a constant $\delta$ as shown in Figure 3.
Thus if $\theta(R)$ is the
angle that the layer normal makes with the tube axis then
$\delta = R\tan\theta(R)$ so that $R$ and $\theta(R)$ obey
\eqn\etanitanii{{\tan\theta(R_1)\over \tan\theta(R_2)} = {R_2\over R_1}.}
Similarly,
the layer spacing is related to $\theta(R)$ by
\eqn\eaiaii{{a(R_1)\over a(R_2)} = \sqrt{1+\tan^2\theta(R_2)\over
1+\tan^2\theta(R_1)}
\equiv{\cos\theta(R_1)\over\cos\theta(R_2)},}
since $\delta\cos\theta(R)=a(R)$.
As $R_2\rightarrow\infty$ for fixed $R_1$, $\tan\theta(R_2)\rightarrow 0$
and $a(R_2)\rightarrow a$, the equilibrium layer spacing
(as it always is for even {\sl single} screw
dislocations).  Both \etanitanii\ and \eaiaii\ come from the equation of the
helicoidal surface.  In one sweep around the azimuthal direction, the surface,
given
in cylindrical co\"ordinates
by the height function $h(r,\phi)=\alpha\phi$, rises by $2\pi\alpha$, {\sl
independent}
of the radius.  This means that the layer normal changes its angle according
to \etanitanii\ and the layer spacing according to \eaiaii.

The second issue is the periodicity of the structure because
it is a cylinder: at radius $R$ the structure must be
$2\pi R$ periodic.  In particular,
when wrapping smectic layers and grain boundaries around a double-twist
cylinder, one must constrain the number of layers or defects to
be integral.  At a given radius $R$, assume that the smectic layers
have rotated by $\theta_0$ from the center.  Additionally, at this same
radius add a grain boundary that rotates the layers by $\Delta\theta$
with a dislocation spacing $d(R)$ and with
all defects having Burgers vector $b=a(R)$, where $a(R)$ is the layer spacing.
If there are $n$ smectic layers in the subsequent smectic region and $m$
defects then trigonometry leads to:
\eqna\econs{$$\eqalignno{{a(R)\over 2d(R)} &=
\sin\left({\Delta\theta\over 2}\right)&\econs a \cr
{md(R)\over 2\pi R}&=\cos\left(\theta_0+{\Delta\theta\over 2}\right)&\econs
b\cr
{na(R)\over 2\pi R}&=\sin\left(\theta_0+\Delta\theta\right)&\econs c\cr}$$}

\noindent where the angle $\theta_0+\Delta\theta/2$ in \econs{b}\ arises
because
the screw dislocations lie symmetrically between the two abutting smectic
regions.
Eliminating $R$, $a(R)$ and $d(R)$ from \econs{}\ leads to:
\eqn\esolv{{m\over n} =
1 - {1\over \cos\Delta\theta + \cot\theta_0\sin(\Delta\theta)}
}
The left-hand side of \esolv\ is a rational number $m/n$.  While
there exists an entire discrete class of solutions to this
Diophantine equation, one class
of solutions has $\cos\Delta\theta = p_1/q_1$ and
$\cot\theta_0=p_2/(q_2\sqrt{q_1^2-p_1^2})$
for $p_1,p_2,q_1,q_2$ integers.
Note that if the
centermost region contains a single screw dislocation down the center of the
double-twist tube then, in order to satisfy \esolv, the first
twist-grain-boundary
will occur at some radius $R$ such that $\tan\theta_c=(a/2\pi R)$ is of the
form
$q_2\sqrt{q_1^2-p_1^2}/p_2$.

Having a class of angles I may now construct many double-twist cylinders that
have the layer normal at the center parallel to the cylinder and the layer
normal
at the edge of the cylinder rotated by $\pi/4$ from this direction.
These are the basic constituents of the blue phases.
At each grain boundary $\theta_0$ in \econs{b,c}\ and \esolv\ is the
sum of the center angle of rotation $\theta_c$ and the sum of all the
consecutive
rotation angles, so if the $i$th boundary is at a radius $R=\rho_i$, then
$\theta_{0,i} = \theta_c +
\sum_{j=1}^{i-1}\tan^{-1}[(\rho_j/\rho_{j+1})\tan(\Delta\theta_j)]$.
Taking into account the natural straightening of the layer normals \etanitanii\
I
can choose radii and angles so that at some
final radius $\rho$ the layer normal can have the desired direction.
Because the tangent angle as
a function of radius is also fixed by the geometry of the screw dislocation
({\sl i.e.},
$\tan[\theta_c(R)] = b/[2\pi R]$) \econs{}\ may be more difficult, if not
impossible, to satisfy if there is a screw dislocation running up the center
of the double-twist tube.

As a simple illustration, let me construct a double-twist tube with layer
normal
$\pi/4$ at some ultimate radius $\rho$ via two grain boundaries both with
a large rotation angle, $\Delta\theta=\sin^{-1}(3/5)$.   This corresponds
to an even simpler class of angles in which both sine and cosine are
rational -- precisely the construction that leads to moir\'e angles.  With
this choice of $\Delta\theta$, I must now choose $\cot\theta_{0,i}$ to
be rational as well.
At the first
boundary (at radius $\rho_1$),
$\theta_{0,1}=0$, so from \esolv\ the ratio of the number of layers to the
number of screw
dislocations is $n_1/m_1=1$, while \econs{a}\ gives
$a(\rho_1)=d(\rho_1)/\sqrt{10}$.
At radius $\rho_2=2\rho_1$, \etanitanii\ gives $\tan\theta_{0,2} = {1\over
2}\tan(\Delta\theta)
= 3/8$ and is rational, as required.  Using the same angle $\Delta\theta$
across
the grain boundary now gives $n_2/m_2 = 12/7 $ and a total tangent of
$\tan(\theta_{0,2}+\Delta\theta) = 36/23$.  Finally, then, at a radius $\rho =
{36\over
23}\rho_2=
{72\over 23}\rho_1$ the layer normal is at an angle $\pi/4$ from the tube
radius.
While this example is rather extreme, involving large angles grain boundaries
and large radii, it is obvious that more realistic examples could
be constructed.

\newsec{Energetics}
Let me now consider the energetics of a smectic double-twist tube.  In
non-layered chiral liquid crystals, double-twist structures are favored
over standard planar cholesteric structures due to a trade-off between the
saddle splay
term \SWM\
\eqn\ess{F_{\rm ss} = \int d^3\!x\,K_{24}\bold{\nabla}\dot\left[\left({\bf\hat
n}\dot\bold{\nabla}\right){\bf\hat n}
- {\bf\hat n}\left(\bold{\nabla}\dot{\bf\hat n}\right)\right],}
and the twist term
\eqn\ess{F_{\rm tw} = \int d^3\!x\,K_2\left[{\bf\hat
n}\dot\left(\bold{\nabla}\cross
{\bf\hat n}\right)
-q_0\right]^2.}
In cylindrical co\"ordinates,
the double-twist texture ${\bf\hat n} = {\bold{\hat z}}\cos(q_0\rho/2) +
{\bold{\hat\phi}}\sin(q_0\rho/2)$.
This director configuration has non-zero saddle splay.  This would be
irrelevant if there were no director defects in the blue phase: $F_{\rm ss}$ is
the integral of a total derivative.  Thus saddle splay is gained at the
surfaces of the cores of nematic defects.  Since the value of the
saddle splay
$2\left[\partial_xn_y\partial_yn_x-\partial_xn_x\partial_yn_y\right]
= -q_0\sin(q_0R)/(2R)$ is non-zero it
can lower the energy compared to the planar texture if $K_{24}>0$.  However,
there is an
energy cost which comes from the
fact that ${\bf\hat n}\dot\left(\bold{\nabla}\cross
{\bf\hat n}\right) = q_0/2 + \sin(q_0R)/(2R)$ and thus the twist
term only vanishes near the center of the double-twist tube, whereas it
vanishes everywhere in the planar cholesteric texture.  Note that in one
double-twist
tube that rotates from $0$ to $\pi/4$, the saddle-splay per unit length
is $F_{\rm ss}/L = -\pi q_0^2(\sqrt{2}-1)/\sqrt{2}$.

The same comparison can be made for layered chiral systems.  First the energy
of the grain boundaries in the double-twist structure must be compared to
the same energy in the standard TGB phase.  Neglecting interactions
between screw dislocations, the energies should be comparable.  A detailed
calculation of the interaction energy would be interesting but,
because of the complex three-dimensional geometry of the dislocations,
difficult
(it should be noted that this calculation has not yet been done even for the
standard TGB state). In some sense this problem is identical to the problem
of deciding between the standard cholesteric phase and the blue phase.
The standard TGB phase has a cholesteric-like organization of the
screw dislocations while the phase proposed here has a blue phase
organization of the screw dislocations.  The deciding factor
between a cholesteric and a blue phase is the presence of an
energetic term which favors or disfavors saddle-splay.
Thus, it is natural to ask whether topological effects play any additional role
in
the smectic double-twist
texture.
In the layered system the nematic disclinations stay the same but now there are
a plethora of screw dislocations.  At each screw dislocation the
layers can pick up surface energies as well.  In fact, these surface energies
are completely analogous to the saddle splay: the Gaussian curvature will
contribute
to the energy at the edges which separate the successive, annular smectic
regions.
For a surface given by a height function $h(x,y)$, the integrated Gaussian
curvature $K$ is:
\eqn\egauin{\int \sqrt{g}d^2\!\sigma K = \int dxdy\,
{\partial_x^2h\partial_y^2h -
\left(\partial_x\partial_yh\right)^2\over\left[1+(\nabla_\perpp
h)^2\right]^{3/2}},}
where $\sqrt{g}dxdy$ is the area element of the surface.
If the helicoidal surface of each smectic layer in region $i$
is given by the multi-valued height function
$h_i(x,y) = \alpha_i\tan^{-1}(y/x)$, then for the region between $\rho_i$ and
$\rho_{i+1}$
\eqn\egauss{\int \sqrt{g_i}d^2\!x K_i =
-2\pi N_i\left[\cos\theta_{0,i+1} -
\cos\left(\theta_{0,i}+\Delta\theta_i\right)\right],}
where $N_i$ is the number of helical turns of the $i$th annular region.
If I consider a segment of the double-twist tube of length $L$, then $N_i
a(\rho_{i+1})=
L\cos\theta_{0,i+1}$
and so the total Gaussian curvature per unit length is
\eqn\egauu{{F_{\rm Gauss}\over L} = -2\pi\sum_{i=1}^N{\cos^2\theta_{0,i+1}\over
a(\rho_{i+1})}\left\{1-{a^2(\rho_i)\over a^2(\rho_{i+1})}\right\}.}
Thus the total Gaussian curvature is always negative (which is no surprise: all
helicoidal
surfaces have negative Gaussian curvature).
The smectic free energy will increase or decrease
depending on whether
the material in question favors or disfavors positive Gaussian curvature,
respectively.
Obviously the energetics will depend on the particular choice of angles
$\Delta\theta_i$
and radii $\rho_i$.  Note that this extra geometric energy of the
double-twisted
smectic is
analogous to the saddle-splay energy of standard blue phases.  The
energy gain from the surface Gaussian curvature
must offset the energy cost of the extra distortions of
the smectic layers and the cholesteric order for double-twist to be favored.
If the net energy change is negative the smectic
will prefer to have double-twist regions instead of the standard TGB
state.

\newsec{Space Filling Structures: Blue Phases}
The ability to
connect the double-twist tubes into one of the standard blue phases
depends on whether the smectic order is short- or long-ranged.
In the first case, since the $N_L^*$ phase is identical to a cholesteric, there
is no equilibrium impediment to constructing the usual blue phase textures.
Since
there is only short range smectic order the integrity of the smectic layers
need
not be globally preserved and there can be any number of edges and screws
filling
the space between the traditional double-twist tubes.  Of course, same-sign
defects repel each other and so there may be dynamical reasons that a complete
blue phase could not form.  It is likely that in the work of \Pansu , that
this melted defect lattice is what is seen.
If a material has a strong tendency to have smectic order then it is reasonable
to
believe that short-range smectic order appears even in the normal blue phases.

The case of true, long-range smectic order is problematic.  Here we
will only consider the $O^5$ blue phase structure
though this work could be extended to the other
known arrangements of double-twist tubes \SWM . The $O^5$ blue phase
has the simple ``log-cabin'' arrangement of double-twist tubes as
shown in Figure 4.  If I arrange for a layer on cylinder $A$ to meet a
layer on cylinder $B$ at the intersection point $\Gamma$ then, since the layer
normal
is at $\pi/4$ with respect to the cylinder axis, that same layer will shift by
$(2\pi)\rho/4$
towards $\Delta$.  Since $\Delta$ is $2\rho$ further along axis $A$ from
$\Gamma$, it
would be impossible for the meeting layer at $\Gamma$ to meet the layer at
$\Delta$.
In Figure 5 I have unwrapped a double-twist tube at the radius $\rho$.  One can
see that
if the layer spacing is $a=a(\rho)$ then for each integer $j$ there is a
layer at $ja(\rho)\sqrt{2}
+\pi\rho/2$ along the (dashed) line on which $\Delta$ lies.
Thus a layer in cylinder $A$ can meet a layer in cylinder $C$ at $\Delta$
if $ja(\rho)\sqrt{2}=(2-\pi/2)\rho$.  At the last grain boundary, however,
$na(\rho_N)=\sin(\theta_{0,N}+\Delta\theta_N)2\pi\rho_N$.
Using \etanitanii\ and \eaiaii\ I find $j/n = (4-\pi)/(4\pi)\approx 1.4165$.
Since
$j/n$ is rational, however, the kissing conditions can never be satisfied.
However,
it would still be possible, though energetically unfavorable, to change
the layer spacing from its preferred value near the cylinder surface.  In this
case some ratio of integers could be found that makes the layer spacing
mismatch small.  Even if this deformation occurred the region between the
double-twist
tubes would have a complicated complexion of screw and edge dislocations
in order to match up {\sl all} the smectic layers.

With some experimental input, it may be possible to show that the
constraints presented in this letter could almost be satisfied for an observed
blue phase with smectic order.  It is most likely that
if a blue phase texture with smectic order persists it would only
have short-range smectic order.  The absence of long-range smectic
order would facilitate a number of the obstacles to the construction of
the double-twist tubes.  For instance, the regions between the double-twist
tubes could melt
and thus alleviate the rational/irrational mismatch.  However,
the special lock-in angles would still correspond to low energy grain
boundaries
since at these angles the smectic order is not frustrated by the
double-twist texture of the layer normal.  I would thus expect that the lock-in
angles would
be present in any blue phase with long- or short-ranged smectic order.

\newsec{Acknowledgments}
It is a pleasure to acknowledge stimulating discussions with
T.C.~Lubensky, B.~Pansu, T.~Powers, J.~Sethna, H.~Stark and an
anonymous referee.
I wish to thank the Aspen Center for Theoretical Physics,
where some of this work was done.
This work was supported by NSF Grant DMR94-23114.

\footatend\vfill\supereject\immediate\closeout\rfile\writestoppt
\baselineskip=.33truein\centerline{{\bf References}}\bigskip{\frenchspacing%
\parindent=20pt\escapechar=` \input refs.tmp\vfill\eject}\nonfrenchspacing
\vfill\eject\immediate\closeout\ffile{\parindent40pt
\baselineskip.33truein\centerline{{\bf Figure Captions}}\nobreak\medskip
\escapechar=` \input figs.tmp\vfill\eject}

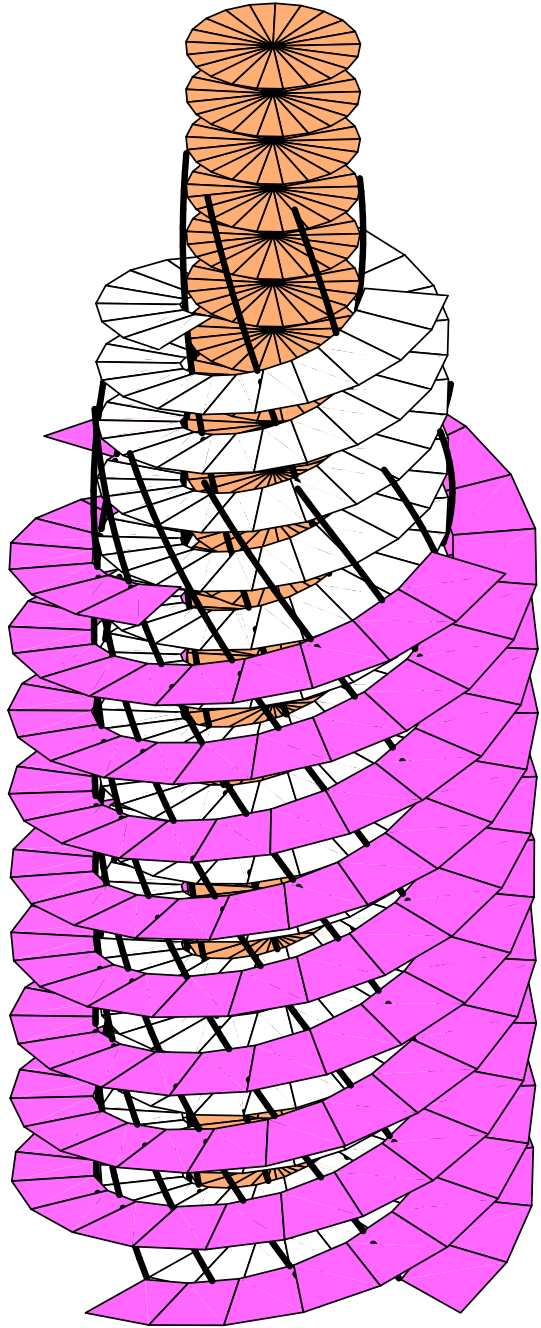
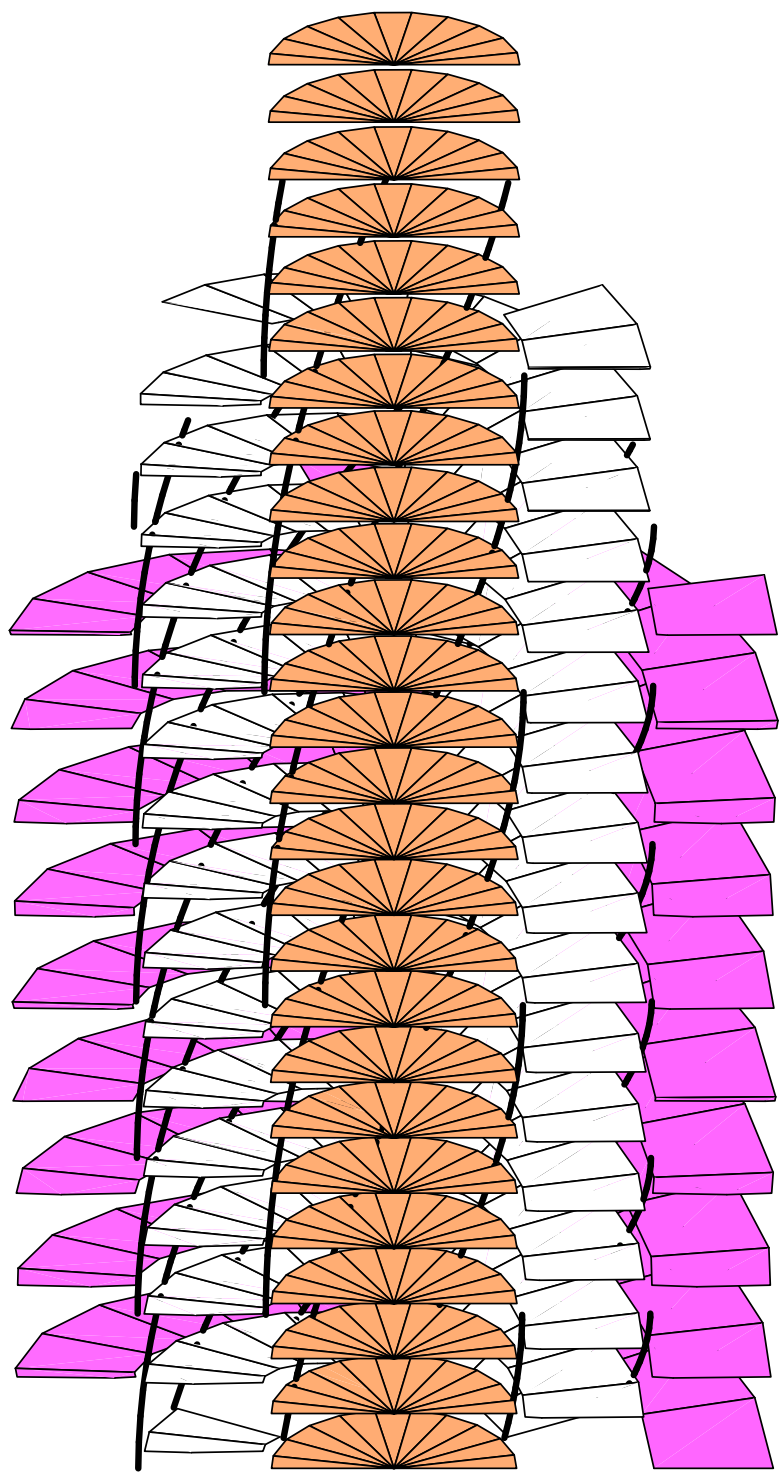
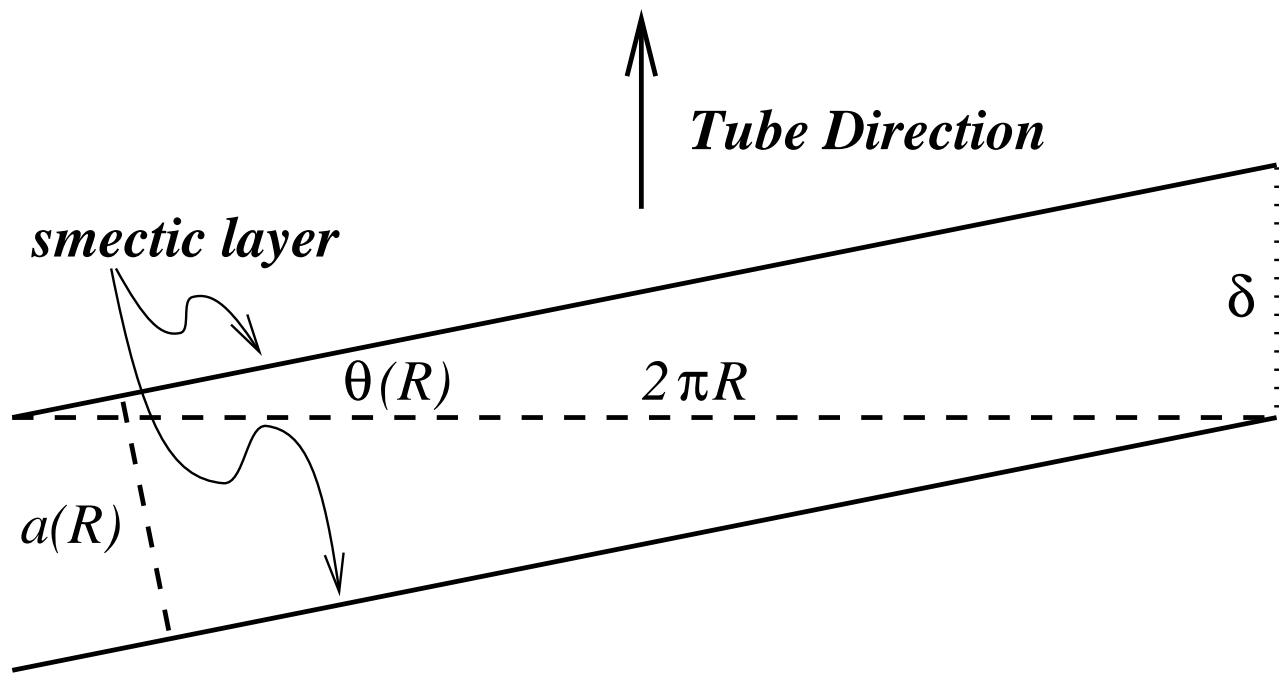
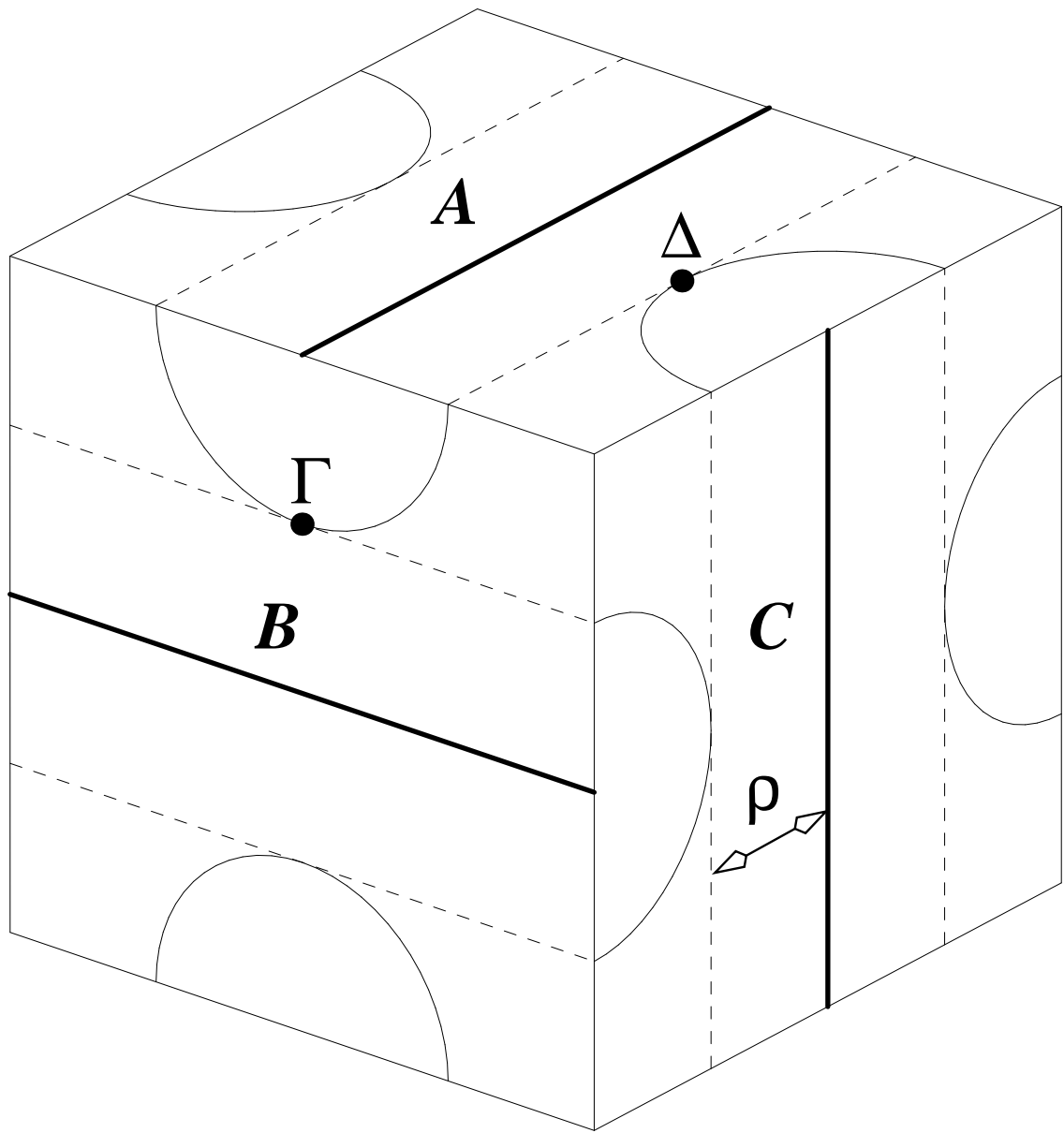
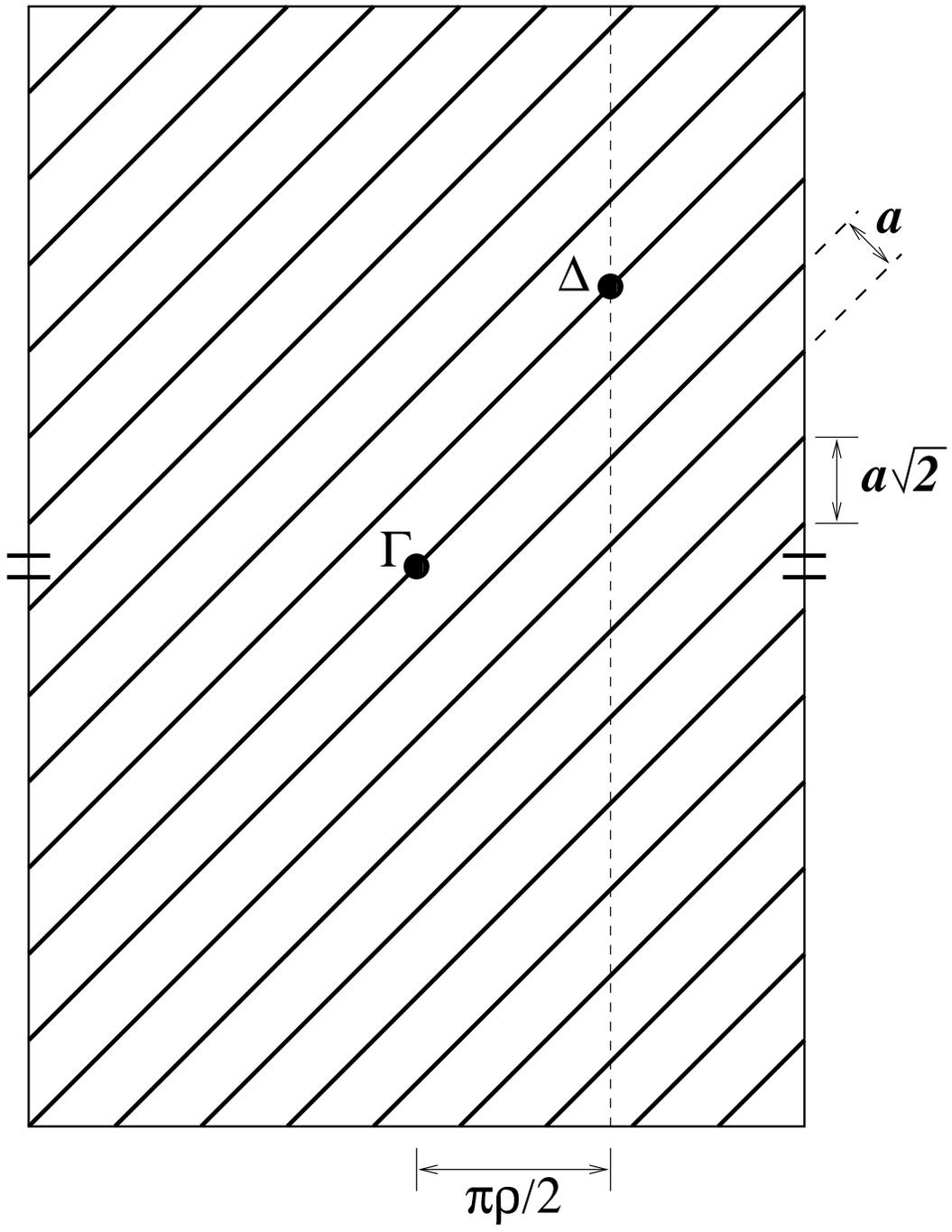
\bye